\documentclass[twocolumn]{aastex62}

\usepackage{apjfonts}

\shorttitle{Central Star of Pa 27}
\shortauthors{Bond \& Zeimann}

\newcommand{\Teff}{T_{\rm eff}}

\newcommand{\kms}{{\>\rm km\>s^{-1}}}

\def\caii{\ion{Ca}{2}}

\def\nai{\ion{Na}{1}}

\def\oiii{\ion{O}{3}}

\def\Gaia{{\it Gaia}}
\newcommand{\GALEX}{{\it GALEX}}

\def\WISE{{\it WISE}}

\newcommand{\TESS}{{\it TESS}}

\begin{document}

\title{Spectroscopic Survey of Faint Planetary-Nebula Nuclei. IV\null. The Abell 35-Type Central Star of Pa~27\footnote{Based on observations obtained with the Hobby-Eberly Telescope (HET), which is a joint project of the University of Texas at Austin, the Pennsylvania State University, Ludwig-Maximillians-Universit\"at M\"unchen, and Georg-August Universit\"at G\"ottingen. The HET is named in honor of its principal benefactors, William P. Hobby and Robert E. Eberly.} }

\author[0000-0003-1377-7145]{Howard E. Bond}
\affil{Department of Astronomy \& Astrophysics, Pennsylvania State University, University Park, PA 16802, USA}
\affil{Space Telescope Science Institute, 
3700 San Martin Dr.,
Baltimore, MD 21218, USA}

\author[0000-0003-2307-0629]{Gregory R. Zeimann}
\affil{Hobby-Eberly Telescope, University of Texas at Austin, Austin, TX 78712, USA}

\correspondingauthor{Howard E. Bond}
\email{heb11@psu.edu}

\begin{abstract}

We present optical spectroscopy of the 12th-mag central star of the planetary nebula (PN) Patchick~27 (Pa~27), obtained during a survey of faint PN nuclei (PNNi) with the Low-Resolution Spectrograph (LRS2) of the Hobby-Eberly Telescope. The optical spectrum of Pa~27 is that of a K0~III red giant with rotationally broadened lines. However, the star is detected in the near-ultraviolet (near-UV) with \GALEX, showing that a hot binary component is also present. The spectral-energy distribution from the near-UV to the mid-infrared can be fitted with a combination of the K0~III giant and a hot PNN with an effective temperature of about 50,000~K. Photometric observations of Pa~27, both ground-based and from \TESS, show a low-amplitude sinusoidal variation with a period of 7.36~days, probably due to starspots on a rotating and magnetically active cool giant. Pa~27 is a new member of the rare class of ``Abell~35-type central stars,'' which are binary PNNi consisting of a spotted late-type star and a hot pre-white dwarf. They are likely the result of a situation where an AGB star ejects its outer layers in a dense wind, part of which is captured by a distant companion, spinning up its rotation by accretion of material and angular momentum. We suggest several useful follow-up observations.

\null\vskip 0.3in

\end{abstract}



\section{Introduction \label{sec:intro} }

The central stars of planetary nebulae (PNe)---planetary-nebula nuclei (PNNi)---exhibit a diverse range of spectroscopic and photometric phenomena, including unusual chemical compositions, binarity, pulsations, strong stellar winds, and rapid evolution { (see review articles by \citealt{Weidmann2020} and \citealt{Kwitter2022}).}
This is the fourth in a series of papers presenting results from a spectroscopic survey of central stars of faint Galactic PNe. The survey is carried out with the second-generation Low-Resolution Spectrograph (LRS2; \citealt{Chonis2016}) of the 10-m Hobby-Eberly Telescope (HET; \citealt{Ramsey1998,Hill2021}), located at McDonald Observatory in west Texas, USA\null. 

An overview of the survey, a description of the instrumentation and data-reduction procedures, target selection, and some initial results, were presented in our first paper \citep[][hereafter Paper~I]{Bond2023a}. Paper~II \citep{Bond2023b} discussed the central star of the ``PN mimic'' Fr~2-30, and Paper~III \citep{WernerPaperIII2024} presents discoveries of three new extremely hot hydrogen-deficient PNNi. 
About 50 central stars have been observed to date. Future papers will discuss several more individual objects of special interest, and another publication will present results on a group of nuclei with fairly normal hydrogen-rich spectra.
In this fourth paper we describe our discovery of a remarkable central star, that of the PN Patchick~27 (Pa~27). It is a new member of a small group of binary nuclei composed of a rotating and spotted late-type star and a hot companion. We present some initial findings, and we encourage follow-up observations.

\section{The Faint Planetary Nebula P\MakeLowercase{a} 27 \label{sec:pa27} }

As explained in our previous papers, we assembled a lengthy target list of central stars for our survey, most of them belonging to faint PNe discovered in recent years by amateur astronomers, and this list was submitted to the HET observing queue (see \citealt{Shetrone2007PASP}). Exposures were chosen for execution by the telescope schedulers, essentially at random, depending on sky conditions and lack of observable higher-ranked targets. 

Pa~27 (PN~G075.0$-$07.2) is a very low-surface-brightness PN discovered by amateur Dana Patchick,\footnote{See \url{https://www.astrobin.com/users/DanaPatchick}} a member of the Deep Sky Hunters (DSH) collaboration. It was included in a list of PNe found by the DSH team that was presented in a conference poster\footnote{Available at \url{http://www.astroscu.unam.mx/apn6/PROCEEDINGS/B3-Kronberger.pdf}} by \citet{Kronberger2014}. The poster shows a deep narrow-band image of Pa~27, obtained with the Kitt Peak 2.1-m telescope in H$\alpha$ and [\oiii] 5007~\AA\null. Pa~27 lies just off the northern edge of the Veil Nebula supernova remnant in Cygnus,\footnote{Deep, wide-angle images of the field are available at \url{https://www.astrobin.com/full/g3myxl/0/} (Andrzej Polkowski) and \url{http://cosmicneighbors.net/pa27-big.jpg} (Mike Keith).} but is an unrelated background object. The PN is roughly circular, with a diameter of about $72''$. The Kitt Peak image shows bright filaments on its northern rim, possibly due to an interaction with the interstellar medium.\footnote{Alternatively, the filaments on the northern edge of Pa~27 could be superposed foreground features belonging to the Veil Nebula. The fact that the proper motion of the PN central star is not in a northerly direction may support this possibility.}  Further information about Pa~27 is given in the online Hong-Kong/AAO/Strasbourg/H$\alpha$ Planetary Nebulae (HASH) database\footnote{\url{http://hashpn.space/}} \citep{Parker2016, Bojicic2017}, including a spectrum of the PN and direct images at several wavelengths from the ultraviolet (UV) to the mid-infrared. 


\section{Central Star \label{sec:centralstar} }

A conspicuous 12th-mag star lies near the center of Pa~27. An image obtained by the {\it Galaxy Evolution Explorer\/} (\GALEX), which is shown at the HASH website, reveals that this star is bright in the near-UV, making it likely to be the ionizing source and central star of the PN\null. 
Table~\ref{tab:gaiadata} gives information on the star, taken from \Gaia\/ Data Release~3\footnote{\url{https://vizier.cds.unistra.fr/viz-bin/VizieR-3?-source=I/355/gaiadr3}} (DR3; \citealt{Gaia2016, Gaia2023}), including its celestial and Galactic coordinates, parallax and proper motion, magnitude and color, and radial velocity and line broadening.

\begin{deluxetable}{lc}[b]
\tablecaption{\Gaia\/ DR3 Data for Central Star of Pa 27  \label{tab:gaiadata} }
\tablehead{
\colhead{Parameter}
&\colhead{Value}
}
\startdata
RA (J2000) & 20 48 58.358  \\
Dec (J2000) & +32 18 14.73 \\
$l$ [deg] &  75.00 \\
$b$  [deg] &  $-7.20$ \\
Parallax [mas] & $0.686\pm0.011$ \\
$\mu_\alpha$ [mas\,yr$^{-1}$] & $+7.586\pm0.008$ \\
$\mu_\delta$ [mas\,yr$^{-1}$] & $+2.653\pm0.012$ \\
$G$ [mag] &  12.27 \\
$G_{\rm BP}-G_{\rm RP}$ [mag] & $1.24$ \\
Radial velocity [$\kms$] & $-47.36\pm2.31$ \\
Line broadening [$\kms$] & $30.8\pm22.3$  \\
\enddata
\end{deluxetable}

A large majority of PNNi have spectra indicating very high surface temperatures, which is true of all of the objects from our survey described in our first three papers. However, the central star of Pa~27 is unusually red at optical wavelengths, with a \Gaia\/ color index of $G_{\rm BP}-G_{\rm RP}=1.24$ (see Table~\ref{tab:gaiadata}).  Since the nucleus is also a \GALEX\/ source, the system is likely to be a binary consisting of the optical star and a UV-bright hot companion.


The reciprocal of the \Gaia\/ parallax gives a distance to the star (and to the nebula) of $1458\pm23$~pc. Using a Bayesian analysis, \citet{BailerJones2021} refine the estimated geometric distance to $1437_{-24}^{+23}$~pc. At this distance, the foreground interstellar extinction, determined using the online {\tt Stilism} tool\footnote{\url{https://stilism.obspm.fr/}} \citep{Capitanio2017}, is $E(B-V)=0.225\pm0.04$. Based on these values, and an apparent magnitude of $V=12.68$ from the APASS catalog,\footnote{\url{https://www.aavso.org/download-apass-data}} we find an absolute magnitude for the central star of $M_V=+1.2$. Thus the optical star appears to be a late-type giant or bright subgiant.


\section{HET/LRS2 Observations and Data Reduction \label{sec:observations} }

Our previous papers give full details of the LRS2 instrumentation used for our survey. In summary, LRS2 is composed of two integral-field-unit (IFU) spectrographs: blue (LRS2-B) and red (LRS2-R), with fields of view of $6''\times12''$. The observations discussed in this paper were made with LRS2-B, which employs a dichroic beamsplitter to send light simultaneously into two units: the ``UV'' channel (covering 3640--4645~\AA\ at a resolving power of 1910), and the ``Orange'' channel (covering 4635--6950~\AA\ at a resolving power of 1140). The data are initially processed using \texttt{Panacea},{\footnote{\url{https://github.com/grzeimann/Panacea}} which performs bias and flat-field correction, fiber extraction, and wavelength calibration. An absolute-flux calibration comes from default response curves and measures of the telescope mirror illumination, as well as the exposure throughput from guider images.  We then apply \texttt{LRS2Multi}\footnote{\url{https://github.com/grzeimann/LRS2Multi}} to the un-sky-subtracted, flux-calibrated fiber spectra, to perform source extraction using a 2$\arcsec$ radius aperture, and subtraction of the spectrum of the sky and nebula based on an annular aperture surrounding the target. Spectra from multiple exposures are then combined. The final spectrum, from both channels, is resampled to a common linear grid with a 0.7~\AA\ spacing. An observation log for our LRS2-B exposures on Pa~27 is presented in Table~\ref{tab:observations}.

\begin{deluxetable}{lc}[h]
\tablecaption{Log of HET LRS2-B Observations of Pa 27\label{tab:observations} }
\tablehead{
\colhead{Date}
&\colhead{Exposure}\\
\colhead{[YYYY-MM-DD]}
&\colhead{[s]}}
\startdata
2020-08-06 & 90 \\
2022-07-15 & 180 \\
2022-08-16 & 180 \\
\enddata
\end{deluxetable}

\section{Optical Spectrum}

We carried out spectral classification for the central star by comparing its spectrum with those of bright stars with known spectral types. We downloaded their spectra from the MILES website\footnote{The acronym stands for Medium-resolution Isaac Newton Telescope Library of Empirical Spectra. See \url{http://miles.iac.es}} \citep{Falcon2011}, which provides a library of observed spectra with a resolution similar to that of our HET data. Guided by the discussion in the next section, we focused on red giants with spectral types of late~G to early~K\null. A good match to the spectrum of Pa~27 was obtained with the K0~III star $\delta$~Cancri. The MILES site gives this star parameters of $\Teff=4657$~K, $\log g=2.51$, and $\rm[Fe/H]=-0.06$. Its absolute magnitude, based on an apparent magnitude of $V=3.94$ \citep{Argue1963} and a \Gaia\/ parallax of 23.83~mas, is $M_V=+0.8$, very close to the value for Pa~27 (+1.2; Section~\ref{sec:centralstar}).

In Figure~\ref{fig:pa27_spectra} we plot our HET LRS2-B spectrum of the central star of Pa~27, split into two segments: 5200--6600~\AA\ (blue line in top panel) and 3800--5200~\AA\ (blue line in bottom panel). Also plotted (red lines in both panels) is the spectrum of $\delta$~Cnc; it has been scaled to the absolute flux of Pa~27, and to match the line profiles of the HET spectrum we applied a boxcar smoothing of 3~pixels (2.7~\AA).

\begin{figure*}[t]
\centering
\includegraphics[width=5in]{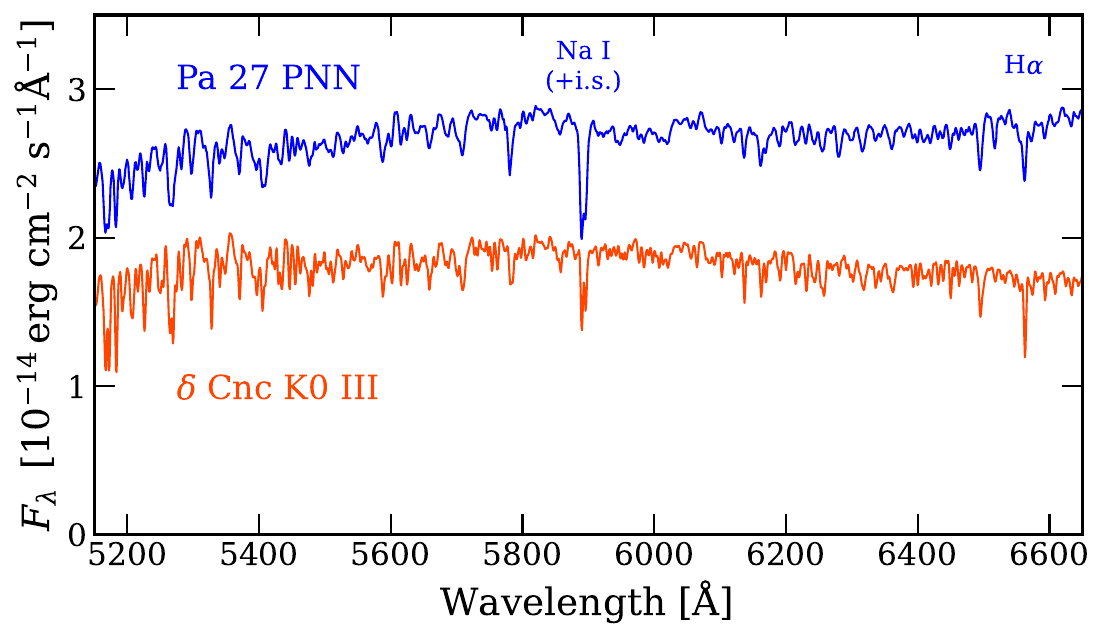}
\includegraphics[width=5in]{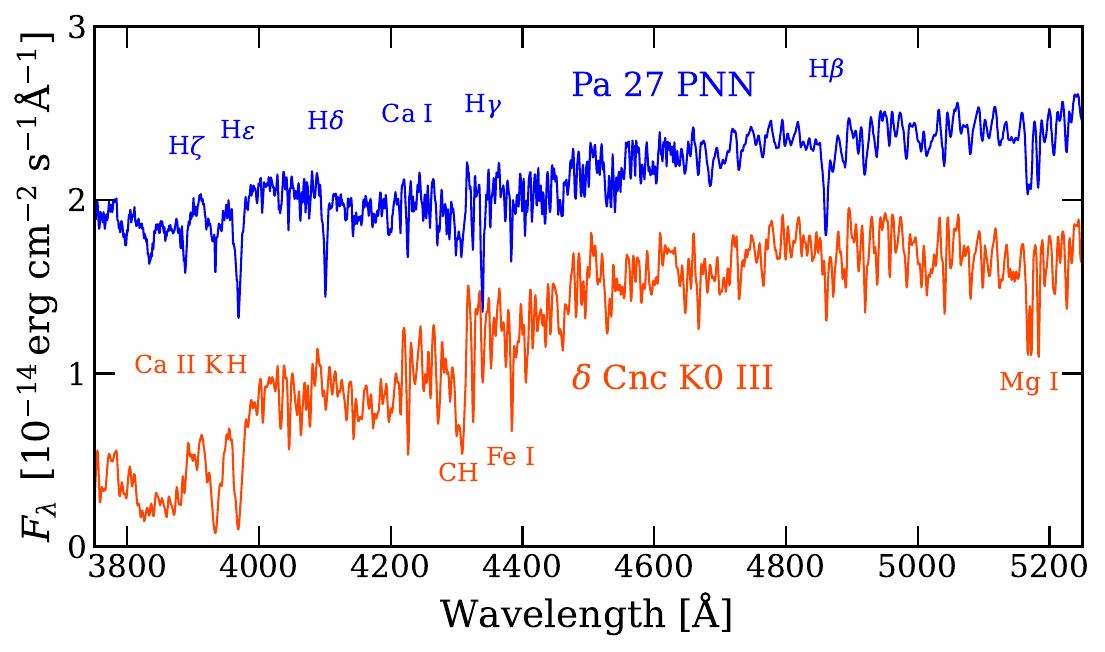}
\caption{
Spectra from 5200 to 6600~\AA\ ({\it top panel}) and 3800 to 5200~\AA\ ({\it bottom panel}). The spectrum of the central star of Pa~27 (from the HET LRS2-B spectrograph) is plotted as a blue line in both panels. For comparison an archival spectrum of the K0 III star $\delta$~Cnc, scaled to the flux level of Pa~27, is plotted as a red line. Several conspicuous spectral features are marked. See text for discussion.
\label{fig:pa27_spectra}
}
\end{figure*}

The top panel in Figure~\ref{fig:pa27_spectra} shows the close similarity of the spectra at longer wavelengths, except that the \nai\ D lines are stronger in Pa~27, due to a component of interstellar absorption.\footnote{One difference is that a line at 6517~\AA\ is stronger in Pa~27 than in $\delta$~Cnc. We suspect that this feature is affected by an instrumental artifact, since a similar artifact is present in several of our spectra of hot central stars. See especially Figure~3 in Paper~I.}

However, in the bottom panel of Figure~\ref{fig:pa27_spectra}, differences in the spectra become more apparent as we move to shorter wavelengths. The atomic lines and CH become weaker in the central star than in $\delta$~Cnc, while the Balmer lines become stronger. The \caii\ K absorption line of the cool star is partially filled in (but note a sharp interstellar component at its core). And the energy distribution of the nucleus becomes shallower than in the K0~III standard star. All of these spectral features indicate the presence of a hot companion of the central star, contributing an increasing amount of flux at shorter wavelengths, and diluting the contribution from the K0 component.

Our IFU frames of the central star show a stellar profile, as do images in publicly available sky surveys, so the binary pair is unresolved in ground-based data. This does not provide a tight constraint on the orbital separation; for example, a separation of $0\farcs5$ would correspond to a projected physical distance as large as $\sim$700~AU.


\section{Spectral-Energy Distribution \label{sec:SED} }

In this section we construct the spectral-energy distribution (SED) of the Pa~27 nucleus. Table~\ref{tab:pa27_magnitudes} presents stellar magnitudes for the central star in a range of bandpasses, and the corresponding absolute fluxes. We gathered these data from the following sources: (1)~A near-UV magnitude from the \GALEX\/ source catalog\footnote{\url{https://galex.stsci.edu/gr6/?page=mastform}} (unfortunately there was no \GALEX\/ observation in the far-UV). (2)~Magnitudes in the $g$, $r$, $i$, $z$, and $y$ bands from the photometric catalog\footnote{\url{https://catalogs.mast.stsci.edu/panstarrs/}} of the Panoramic Survey Telescope and Rapid Response System (Pan-STARRS or PS1; \citealt{Kaiser2010}). (3)~Near-infrared magnitudes from the Two Micron All Sky Survey (2MASS; \citealt{Skrutskie2006}). (4)~Mid-infrared magnitudes from the {\it Wide-field Infrared Survey Explorer\/} \citep[\WISE;][]{Wright2010}. The 2MASS and \WISE\/ data were obtained from the NASA/IPAC Infrared Science Archive.\footnote{\url{http://irsa.ipac.caltech.edu/frontpage}}


\begin{deluxetable}{lcccc}[h]
\tablecaption{Spectral-Energy Distribution of Pa 27 Central Star \label{tab:pa27_magnitudes} } 
\tablewidth{0pt}
\tablehead{
\colhead{Bandpass}  &
\colhead{Magnitude\tablenotemark{a}}  &
\colhead{Source\tablenotemark{a}}  &
\colhead{$\lambda_{\rm eff}$\tablenotemark{b}}  &
\colhead{$F_\lambda$\tablenotemark{c}}  \\
\colhead{ }  &
\colhead{ }  &
\colhead{ }  &
\colhead{[$\mu$m]}  &
\colhead{[$\rm erg\,cm^{-2}\,s^{-1}\,$\AA$^{-1}$]}  
}  
\startdata 
NUV   & 14.467  & \GALEX\/    & 0.2267 & $3.62\times10^{-14}$  \\
\noalign{\vskip0.1in} 
$g$   & 13.163 & PS1 	  & 0.481 & $2.56\times10^{-14}$  \\
$r$   & 12.295 & PS1 	  & 0.617 & $3.45\times10^{-14}$  \\
$i$   & 11.951 & PS1 	  & 0.752 & $3.19\times10^{-14}$  \\
$z$   & 11.762 & PS1 	  & 0.866 & $2.86\times10^{-14}$  \\
$y$   & 12.637 & PS1 	  & 0.962 & $2.60\times10^{-14}$  \\
\noalign{\vskip0.1in} 
$J$   & 10.543 & 2MASS   & 1.235  & $1.90\times10^{-14}$  \\
$H$   & 9.918 & 2MASS    & 1.662  & $1.22\times10^{-14}$  \\
$K_s$ & 9.756 & 2MASS    & 2.159  & $5.36\times10^{-15}$  \\
\noalign{\vskip0.1in} 
$W$1  & 9.633 & \WISE\/  & 3.35   & $1.16\times10^{-15}$  \\
$W$2  & 9.668 & \WISE\/  & 4.60   & $3.30\times10^{-16}$  \\
$W$3  & 9.339 & \WISE\/  & 11.56  & $1.31\times10^{-17}$  \\
$W$4  & 5.785 & \WISE\/  & 22.09  & $2.49\times10^{-17}$  \\
\enddata 
\tablenotetext{a}{Sources for magnitudes in column 2 are given in column~3 and described in the text. Uncertainties are about $\pm$0.01~mag for \GALEX, $\pm$0.01--0.02~mag for PS1, $\pm$0.02~mag for 2MASS, and $\pm$0.02--0.03~mag for \WISE\/ $W1$ to $W3$. The $W4$ magnitude is heavily contaminated by the surrounding PN. }
\tablenotetext{b}{Effective wavelengths are from the following sources. \GALEX\/ NUV: \url{https://asd.gsfc.nasa.gov/archive/galex/}; PS1: \citet{Tonry2012}; 2MASS: \url{http://coolwiki.ipac.caltech.edu/index.php/Central_wavelengths_and_zero_points}; \WISE: \citet{Wright2010}.}
\tablenotetext{c}{Absolute fluxes were determined using photometric zero-points for 2MASS and \WISE\/ from the Caltech compilation cited in Footnote~b. \GALEX\/ and PS1 magnitudes are on the AB scale.}. 
\end{deluxetable}

The final column in Table~\ref{tab:pa27_magnitudes} gives absolute fluxes for the central star, converted from the magnitudes in column~2 using the zero-points referenced in the table footnotes. These fluxes are plotted against wavelength as filled circles in Figure~\ref{fig:pa27_sed}, color-coded as indicated in the figure legend. We also plot our calibrated HET spectrum as an orange line. 

\begin{figure}[h]
\centering
\includegraphics[width=0.47\textwidth]{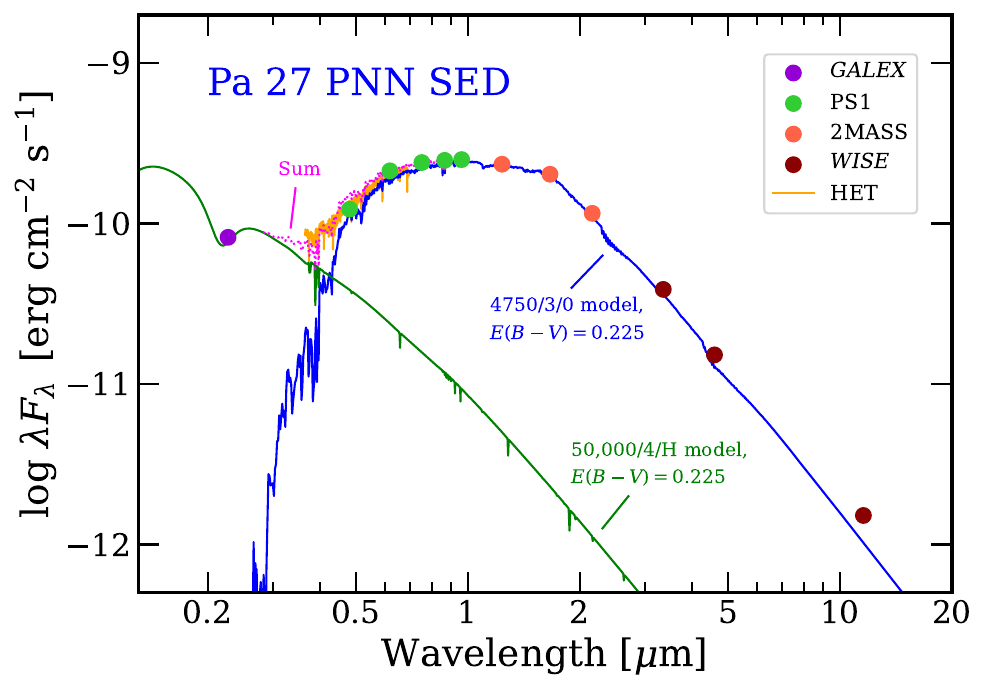}
\caption{
Spectral-energy distribution for the nucleus of Pa~27. Photometric fluxes from \GALEX, Pan-STARRS, 2MASS, and \WISE\/ are plotted as filled circles, color-coded as indicated in the legend. The orange line is our calibrated HET spectrum. The SED indicates that two stars are present. The blue curve represents a model atmosphere for a star with $\Teff=4750$~K, $\log g=3$, and solar metallicity, with its flux scaled to match the observed spectrum longward of $\sim$7000~\AA\null. The green curve is for a pure-hydrogen atmosphere with $\Teff=50,000$~K and $\log g=4$, scaled to match the \GALEX\/ near-UV flux. The dotted magenta curve is the sum of the two models. A reddening of $E(B-V)=0.225$ has been applied to both model spectra. See text for further discussion.
\label{fig:pa27_sed}}
\end{figure}

The SED clearly shows an excess in the near-UV, confirming the presence of two stars in the system. In order to estimate the parameters of the cool component, we compared the SED with synthetic model-atmosphere spectra computed using {\tt ATLAS9} \citep{Castelli2003}, which are conveniently available online at the Spanish Virtual Observatory (SVO) website.\footnote{\url{http://svo2.cab.inta-csic.es/theory/newov2}} Based on the known absolute magnitude, we assumed $\log g=3$. We applied a reddening of $E(B-V)=0.225$ (see Section~\ref{sec:centralstar}) to each theoretical spectrum, using the formulation of \citet{Cardelli1989}, and then scaled it to the observed flux level. The best fit to the SED points longward of about 7000~\AA\ was obtained for a model with parameters $\Teff=4750$~K, $\log g=3$, and $\rm[Fe/H]=0$. These are consistent with the K0~III spectral type derived in the previous section. The SED of this reddened theoretical model is plotted as a blue line in Figure~\ref{fig:pa27_sed}. 

The $\sim$4750~K cool component provides a very good { match} to the SED points at the $i$ band and longward, but the hot companion contributes increasing amounts of flux at shorter wavelengths. Unfortunately, the parameters of the hot star can only be loosely constrained, since we have just the single \GALEX\/ point in the near-UV, along with a small blue excess seen at optical wavelengths. We selected several theoretical SEDs for hot stars computed with the T\"ubingen Model-Atmosphere Package \citep[TMAP;][]{Werner2003}, which are available at the SVO website cited above. Given the strong Balmer lines in the optical spectrum, we considered pure-hydrogen models, and we assumed $\log g=4$ based on typical post-asymptotic-giant-branch (post-AGB) evolutionary tracks { (see, e.g., Figure~7 in our Paper II).} We then applied a reddening of $E(B-V)=0.225$, and scaled each TMAP SED to match the \GALEX\/ near-UV point. We obtained a reasonable { match} with a $\Teff=50,000$~K model, shown as the green line in Figure~\ref{fig:pa27_sed}. The magenta dotted line shows the sum of the two theoretical SEDs. { This sum} provides a good {match} to the HET spectrum and the photometric points shortward of the $i$ band, and explains the dilution of the K0 spectrum in the bottom panel of Figure~\ref{fig:pa27_spectra}. However, our adopted { nominal} temperature for the hot component must be considered as only approximate.

\section{Photometric Variability}

The nucleus of Pa~27 was found to be a periodic variable star by the Asteroid Terrestrial-impact Last Alert System \citep[ATLAS;][]{Heinze2018}, designated ATO J312.2432+32.3041. It is described\footnote{See the ATLAS variable-star catalog at \url{https://vizier.cds.unistra.fr/viz-bin/VizieR?-source=J/AJ/156/241}} as having sine-wave variability with a period of 7.365430~days and a peak-to-peak range of 0.09~mag in a ``cyan'' filter and 0.10~mag in an ``orange'' filter. Very similar results are given by the All-Sky Automated Survey for Supernovae \citep[ASAS-SN;][]{Shappee2014, Kochanek2017}, whose catalog\footnote{\url{https://asas-sn.osu.edu}} designates the object as ASASSN-V J204858.33+321814.7. ASAS-SN classifies the star as a ``rotational'' variable, i.e., a rotating spotted star. The period was found to be 7.3677381~days, with a $V$-band peak-to-peak amplitude of 0.10~mag.

We obtained aperture photometry of the central star from the {\it Transiting Exoplanet Survey Satellite\/} (\TESS) mission, using the online {\tt TESSExtractor} tool\footnote{\url{https://www.tessextractor.app}} \citep{Serna2021}. \TESS\/ monitors sky ``Sectors'' for continuous durations of $\sim$27~days.  The location of Pa~27 has been observed by \TESS\/ during the Sector~15 run (30-minute cadence), and the Sectors~41 and 55 runs (10-minute cadences). We edited the {\tt TESSExtractor} downloads to remove data affected by bright background and other instrumental effects.  We determined eight times of maximum light, over an interval of 1097~days,  by fitting parabolas to the light-curve peaks, and calculated a linear fit to these times. The derived ephemeris is 
\begin{eqnarray}
T_{\rm max} \, ({\rm BJD}) \, - 2457000 = (1717.300 \pm 	0.111) \nonumber \\
+ \, (7.3638 \pm	0.0011)\,E \, .\nonumber 
\end{eqnarray} 
The \TESS\/ light curves, phased to this ephemeris, are plotted in Figure~\ref{fig:pa27_lightcurve}. The curves are nearly perfect sine waves, with peak-to-peak amplitudes averaging 0.045~mag in the broadband \TESS\/ photometric system.

\begin{figure}[h]
\centering
\includegraphics[width=0.47\textwidth]{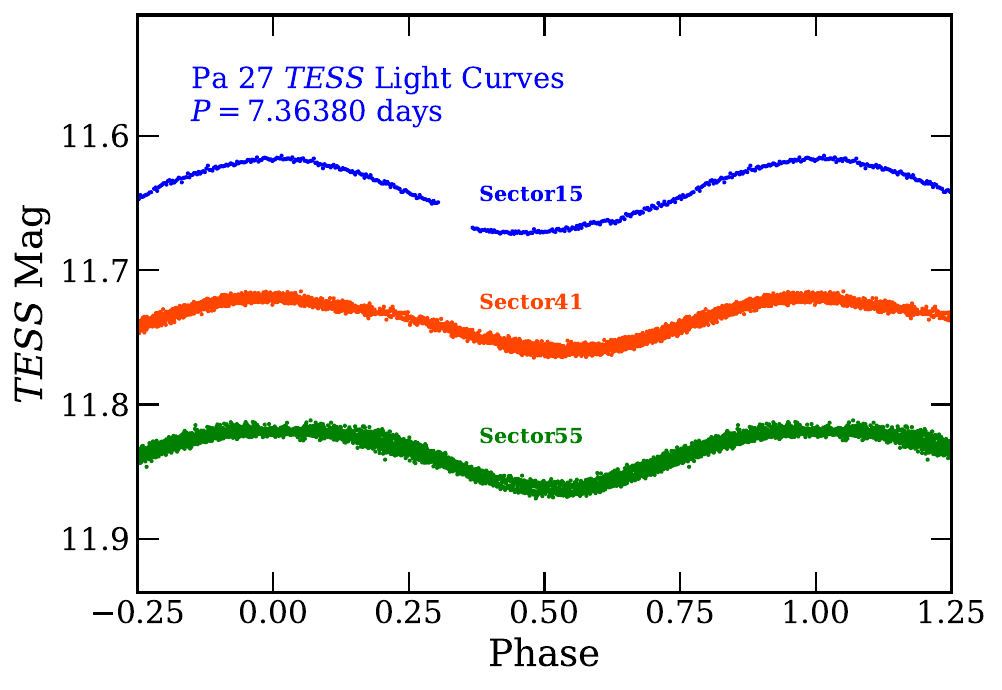}
\caption{
Light curves of Pa~27 from three \TESS\/ runs, phased with the ephemeris given in the text. For clarity, the curves from Sectors 41 and 55 are shifted fainter by 0.1 and 0.2 mag, respectively, relative to the Sector 15 curve. 
\label{fig:pa27_lightcurve}}
\end{figure}

We attribute the variability of Pa~27 to dark starspots on its rotating K-type red giant. In principle, a sinusoidal light curve of the cool component could alternatively be due to a close binary with a heated hemisphere producing one light maximum per orbital period, or to ellipsoidal variations in a system with twice the period given above. However, at least two arguments support the rotational interpretation. (1)~A well-established class of rotationally variable late-type PNNi, with UV-bright hot companions, exists (see next section). (2)~The phases and amplitudes of variation in Pa~27 change with time. First, the times of maximum light in \TESS\/ Sector~41 are slightly earlier in phase than in the other two \TESS\/ runs. Second, the peak-to-peak amplitudes for the three individual high-precision \TESS\/ runs are slightly variable, at 0.056, 0.040, and 0.044~mag, respectively. More dramatically, the amplitudes seen at the epochs of the \TESS\/ observations are less than half that seen at earlier epochs in the orange filter of ATLAS\null. The orange bandpass has a similar effective wavelength to that of \TESS\null. These time-variable behaviors would not occur due to thermal or geometric effects in a close binary, but are consistent with starspots that vary in both location on the star, and in the amount of coverage of the stellar surface. 

{

As a check on this interpretation, we estimated the radius of the star, using an absolute magnitude of $M_V=+1.2$ (Section~\ref{sec:centralstar}) and an {\tt ATLAS9} bolometric correction of 0.48~mag, yielding $R=9.5\,R_\odot$. For a rotation period of 7.3638~days, we find an equatorial rotational velocity of $65\,\kms$. This is consistent with the \Gaia\/ line broadening measurement of $v\sin i\simeq30.8\pm22.3\,\kms$ from Table~\ref{tab:gaiadata}, given that the inclination angle is unknown.

}

\section{P\MakeLowercase{a} 27 as an A\MakeLowercase{bell} 35-type Planetary Nucleus}

To summarize the above results, the central star of the PN Pa~27 is a binary system composed of a K0~III red giant, accompanied by an unresolved UV-bright hot companion with an effective temperature of roughly 50,000~K\null. The starspotted K star has rotationally broadened spectral lines, and is a low-amplitude variable with a photometric period of 7.36~days. 

These findings show that Pa~27 belongs to a small class of PNNi for which the name ``Abell 35-type central stars'' was proposed by \citet{Bond1993}. These objects are binary nuclei of PNe consisting of a rotating and spotted late-type star and a hot companion. \citet{Bond1993} listed the three members of the class known at that time: Abell~35, LoTr~1, and LoTr~5, along with a field star with similar properties, HD~128220. 

A physical explanation for Abell~35-type nuclei was proposed by \citet[][hereafter JS96]{Jeffries1996}: they arise in a situation where an AGB star ejects a dense wind, part of which is captured by a moderately distant companion star. The companion accretes material and angular momentum from the wind, spinning up its rotation. The remnant core of the AGB star is now an UV-bright hot white dwarf or pre-white dwarf. JS96 describe the spun-up cool companions as ``wind-accretion induced, rapidly rotating stars (or WIRRing stars).'' Their calculations suggest that the spin-ups can occur in binaries with separations as large as $\sim$100~AU\null. 

In this picture, rotation drives magnetic surface activity on the late-type star, creating extensive starspots and hence photometric variations at the spin period. The periods of the light variations are much shorter than the orbital periods of the binaries. 


The prototypical system, Abell~35 itself, was found to have a nucleus\footnote{Several authors \citep[e.g.,][]{FrewParker2010, Ziegler2012} have argued that the Abell~35 nebula is actually not a true PN, but is instead a ``PN mimic,'' created when a hot star passes through and photoionizes an overdense region of the interstellar medium.} with a late-type (G8~III-IV) spectrum by \citet{Jacoby1981}. Periodic light variations due to rotation of the star were discovered by \citet{Jasniewicz1988}. UV observations reveal a hot companion \citep[][and references therein]{Herald2002}. To our knowledge, the orbital period of the binary remains unknown, but it is longer than at least several decades \citep[e.g.,][]{Gatti1998}. 

JS96 predicted that accretion of processed material from the AGB wind could create overabundances of carbon and $s$-process elements on the WIRRing companion. This prediction was borne out by the discovery by \citet{BondWeBo12003} of a barium star---a cool red giant with overabundances of carbon and $s$-process elements---at the center of the PN WeBo~1. The nucleus of WeBo~1, at optical wavelengths, is a late-type star with photometric variability at a period of 4.7~days. Its hot companion is detected in the UV \citep{Siegel2012}. Another barium-star PNN, that of Abell~70, was discovered by \citet{Miszalski2012}; it too is a photometric variable, at a period of 2.06~days, with a changing amplitude \citep{BondCiardullo2018,JonesAbell702022}.

The system of LoTr 1 is remarkably similar to Pa~27, with a nearly identical absolute magnitude of $M_V=+1.3$. It was studied in detail by \citet{Tyndall2013}, who show that its central star is a binary containing a K1~III giant and a hot companion, showing photometric variations with a period of 6.4~days. Its orbital period is as yet unknown.
The central star of the PN LoTr~5 is still another case of a rapidly rotating G-type star and a UV-bright companion. It has a photometric period of 5.95~days, but in this case the orbital period of the binary has been established to be about 2700~days \citep[][and references therein]{Aller2018}. Lastly, the central star of the PN Hen 2-39 is a late-type barium star, again with a hot companion \citep{LoblingHen2-392019}. It is a photometric variable with a period of 5.46~days \citep{Miszalski2013}. 

We used {\tt TESSExtractor} to obtain \TESS\/ light curves of the relatively bright central stars of Abell~35, LoTr~1, LoTr~5, and WeBo~1.\footnote{Unfortunately, the central stars of Abell~70 and Hen 2-39 are too faint and/or crowded for useful \TESS\/ photometry.} We downloaded single sector runs for each of these targets, and they are compared with a single run on Pa~27 in Figure~\ref{fig:pa27_lightcurve_collection}. Remarkably, all five objects show almost pure sinusoidal variations, but with a range of timescales from 0.77~days for Abell 35 to 7.36~days for Pa~27. The photometric amplitudes are similar as well.

\begin{figure*}
\centering
\includegraphics[width=5in]{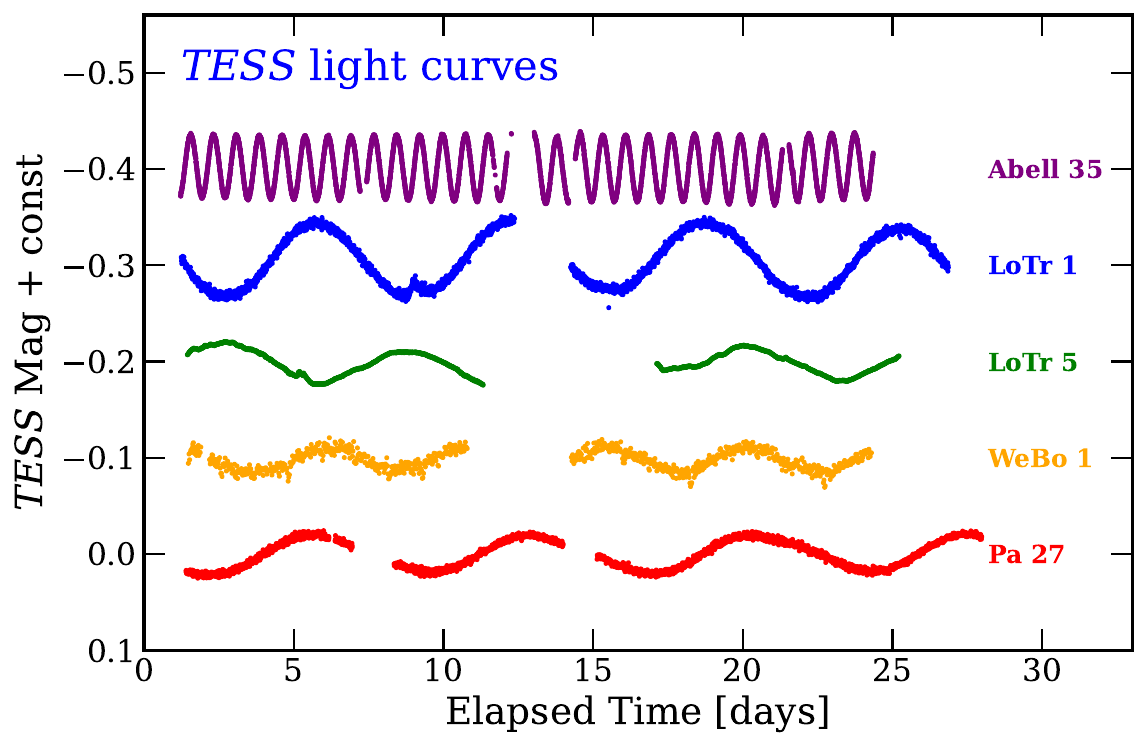}
\caption{
\TESS\/ light curves for central stars of Abell~35 (Sector 64), LoTr~1 (Sector 32), LoTr~5 (Sector 49), and WeBo~1 (Sector 49), compared with the Sector~41 run on Pa~27.  
\label{fig:pa27_lightcurve_collection}}
\end{figure*}

\section{Future Studies}

The main purpose of this paper is to encourage further studies of the central star of Pa~27, a new member of the Abell~35 class of binary systems. Radial-velocity (RV) monitoring could establish the orbital period of the binary, which is likely to be much longer than the 7.36-day spin period of the cool component. Our three HET spectra do not provide useful information on this question. However, \Gaia\/ DR3 indicates a peak-to-peak range of $47.5\,\kms$ in its RV, so binary motion has been detected. Unfortunately, at this writing, the individual \Gaia\/ RVs are not publicly available.

UV spectra could provide constraints on the nature of the hot pre-white-dwarf companion.

An abundance-analysis study of this relatively bright star would be of interest. We see no gross evidence for enhancements of carbon and $s$-process elements like strontium and barium in our spectra, but a study at higher resolution and signal-to-noise would be more informative.


\acknowledgments

We thank the HET queue schedulers and nighttime observers at McDonald Observatory for obtaining the data discussed here.


The Low-Resolution Spectrograph 2 (LRS2) was developed and funded by The University of Texas at Austin McDonald Observatory and Department of Astronomy, and by The Pennsylvania State University. We thank the Leibniz-Institut f\"ur Astrophysik Potsdam (AIP) and the Institut f\"ur Astrophysik G\"ottingen (IAG) for their contributions to the construction of the integral-field units.

We acknowledge the Texas Advanced Computing Center (TACC) at The University of Texas at Austin for providing high-performance computing, visualization, and storage resources that have contributed to the results reported within this paper.

Based on data from the MILES library service developed by the Spanish Virtual Observatory in the framework of the IAU Commission G5 Working Group: Spectral Stellar Libraries.


This work has made use of data from the European Space Agency (ESA) mission
{\it Gaia\/} (\url{https://www.cosmos.esa.int/gaia}), processed by the {\it Gaia\/}
Data Processing and Analysis Consortium (DPAC,
\url{https://www.cosmos.esa.int/web/gaia/dpac/consortium}). Funding for the DPAC
has been provided by national institutions, in particular the institutions
participating in the {\it Gaia\/} Multilateral Agreement.

This work is based in part on observations made with the NASA {\it Galaxy Evolution Explorer}. \GALEX\/ was operated for NASA by the California Institute of Technology under NASA contract NAS5-98034.

The Pan-STARRS1 Surveys (PS1) have been made possible through contributions of the Institute for Astronomy, the University of Hawaii, the Pan-STARRS Project Office, the Max-Planck Society and its participating institutes, the Max Planck Institute for Astronomy, Heidelberg and the Max Planck Institute for Extraterrestrial Physics, Garching, The Johns Hopkins University, Durham University, the University of Edinburgh, Queen's University Belfast, the Harvard-Smithsonian Center for Astrophysics, the Las Cumbres Observatory Global Telescope Network Incorporated, the National Central University of Taiwan, the Space Telescope Science Institute, the National Aeronautics and Space Administration under Grant No.\ NNX08AR22G issued through the Planetary Science Division of the NASA Science Mission Directorate, the National Science Foundation under Grant No.\ AST-1238877, the University of Maryland, and Eotvos Lorand University (ELTE).

This research was made possible through the use of the AAVSO Photometric All-Sky Survey (APASS), funded by the Robert Martin Ayers Sciences Fund and NSF AST-1412587.

This publication makes use of data products from the Two Micron All Sky Survey, which is a joint project of the University of Massachusetts and the Infrared Processing and Analysis Center/California Institute of Technology, funded by the National Aeronautics and Space Administration and the National Science Foundation.

This publication makes use of data products from the Wide-field Infrared Survey Explorer, which is a joint project of the University of California, Los Angeles, and the Jet Propulsion Laboratory/California Institute of Technology, funded by the National Aeronautics and Space Administration.

Funding for the \TESS\/ mission is provided by NASA's Science Mission directorate.

This research has made use of the SIMBAD and Vizier databases, operated at CDS, Strasbourg, France.

This research has made use of the Spanish Virtual Observatory (\url{https://svo.cab.inta-csic.es}) project funded by MCIN/AEI/10.13039\slash 501100011033/ through grant PID2020-112949GB-I00.

\bibliography{PNNisurvey_refs}

\end{document}